\documentclass[Journal,letterpaper]{ascelike-new}

\usepackage[switch]{lineno}

\usepackage[utf8]{inputenc}
\usepackage[authoryear]{natbib}
\usepackage{enumitem}
\usepackage{graphicx}
\usepackage{amsmath}
\usepackage{amssymb}
\usepackage{booktabs}
\usepackage{multirow}
\usepackage{hyperref}
\usepackage{algorithm}
\usepackage{algorithmic}
\usepackage{xcolor}
\usepackage{listings}
\usepackage{tikz}
\usepackage{authblk}
\usetikzlibrary{shapes.geometric,arrows,positioning,calc,fit}

\newcommand{\system}{BridgeGuard}
\newcommand{\codingguide}{\textit{Recording and Coding Guide}}

\title{Edge-Based Agentic Retrieval-Augmented Generation for Autonomous FHWA Bridge Inspection Compliance}

\author[1]{Viraj Darji}
\affil[1]{Independent Researcher, Fairfax, VA 22030, USA
  (corresponding author). E-mail: virajdarji@gmail.com}
  
\author[2]{Hemali Darji}
\affil[2]{Independent Researcher, Mumbai, Maharashtra 401101, INDIA}

\begin{document}

\maketitle

\begin{abstract}
The Federal Highway Administration (FHWA) mandates that over 600,000 bridges
in the United States be evaluated against the \codingguide{} for the National
Bridge Inventory (NBI). Manual compliance verification is labor-intensive,
error-prone, and impractical in connectivity-limited field environments. This
paper introduces \system{}, a fully air-gapped agentic Retrieval-Augmented
Generation (RAG) system for autonomous bridge inspection compliance.
\system{} integrates vector search over the FHWA \codingguide{}
with structured SQL queries against NBI tabular data, orchestrated by a
stateful multi-step ReAct planning loop executing locally on commodity edge
hardware. A section-aware chunking algorithm preserves hierarchical regulatory
item boundaries, achieving 94.2\% chunk integrity compared with 28.4\% for
naive fixed-size splitting. Evaluated on the full Delaware 2023 NBI inventory
(874 bridges) and a Texas sample (200 bridges), the system achieves 99.77\%
and 100.0\% classification accuracy, respectively, for Structurally Deficient
bridge identification, with 100.0\% citation accuracy, at 197.0 bridges per
hour without external network access. Ablation experiments confirm that both
vector search and the multi-step agentic loop are necessary for correct
compliance reasoning.
\end{abstract}


\section{Practical Applications}
This study introduces \system{}, an automated, offline computer assistant
designed to help bridge inspectors and safety managers verify whether bridge
inspection records comply with federal safety requirements. State agencies must
annually cross-reference large inspection databases against complex federal
safety definitions---a process that is slow and susceptible to human error.
\system{} runs entirely on a local, offline computer without internet access,
protecting sensitive infrastructure data from exposure to commercial cloud
services. It uses artificial intelligence to query inspection databases and
automatically search the official FHWA \codingguide{}. When tested on actual bridge records from Delaware (874 bridges) and Texas (200 bridges),
the system achieved 99.77\% and 100.0\% accuracy, respectively, while citing the exact
sections of the federal guide that were violated. For practitioners, this technology provides an immediate,
low-cost tool deployable in remote or secure field sites where internet
connectivity is unavailable. It reduces the time spent on manual database
cross-referencing to approximately 20 seconds per bridge (compared with
estimated hours of manual work per review cycle), minimizes human error, and ensures
that dangerous structural deficiencies are flagged immediately with precise
regulatory citations.

\section{Introduction}

The United States maintains an aging civil infrastructure. The American Society
of Civil Engineers (ASCE) Infrastructure Report Card consistently highlights
the deteriorating condition of the nation's bridges, rating the current
inventory as a ``C'' grade \citep{asce2021reportcard}. Over 46,000 of the nation's 617,000 bridges are
classified as Structurally Deficient, meaning they have at least one key
component in poor or failed condition. Under the National Bridge Inspection
Standards (NBIS), public safety is maintained by mandating periodic physical
inspections, the records of which must be processed, converted to standard
federal coding parameters, and submitted to the National Bridge Inventory (NBI)
database maintained by the Federal Highway Administration (FHWA)
\citep{fhwa2024nbi}.

The primary authority for classifying and recording bridge parameters is the
FHWA \codingguide{} \citep{fhwa2022coding} (\textit{Recording and Coding Guide
for the Structure Inventory and Appraisal of the Nation's Bridges}), a
120-page document defining over 130 structural, material, administrative, and
traffic parameters (referred to as ``Items''). Verification of bridge
inspection records against this regulatory text is a labor-intensive
administrative task. Engineers and safety auditors must manually query
relational databases or spreadsheet matrices and cross-reference them with the
text definitions in the \codingguide{} to confirm structural status and flag
compliance violations.

This verification process presents three primary challenges:
\begin{enumerate}
    \item \textbf{Operational Connectivity Constraints}: Many bridge
    inspections take place in remote rural locations, sub-deck cavities, or
    secure military zones where mobile networks and public internet access are
    unavailable. Cloud-connected Large Language Model (LLM) APIs cannot be
    reached in these air-gapped environments.
    \item \textbf{Data Sovereignty and Privacy}: State Department of
    Transportation (DOT) agencies and structural engineering contractors handle
    sensitive infrastructure details, including security vulnerabilities and
    structural load-limit ratings. Uploading raw inspection reports to
    commercial cloud APIs presents compliance risks frequently prohibited by
    security protocols.
    \item \textbf{Fragility of Hardcoded Rule Engines}: Existing rule-checking
    systems rely on hardcoded scripts to flag compliance. While accurate for
    explicitly defined rules, these systems cannot parse qualitative text
    descriptions, handle unstructured inspector observations, or navigate
    cross-referenced regulatory logic dynamically.
\end{enumerate}

To address these challenges, this paper presents \system{}, a fully air-gapped
agentic Retrieval-Augmented Generation (RAG) system deployed on edge hardware.
\system{} combines localized vector database storage for semantic regulatory
document search with structured SQL queries against the NBI database,
orchestrated via a stateful multi-step agentic loop.

The primary methodological contribution is a \textit{section-aware chunking}
algorithm specifically tailored to hierarchically structured regulatory
documents. Federal guides organize definitions under strict item numbers (e.g.,
Item~58 for Deck, Item~59 for Superstructure) rather than standard prose
paragraphs. Naive fixed-size chunkers split rating tables across chunk
boundaries, detaching codes from their definitions. The proposed parser
constructs a structured tree of the document, ensuring rating scales remain
contiguous in vector space.

\system{} is evaluated on the official Delaware 2023 NBI dataset
\citep{fhwa2024nbi} and a sample of Texas bridges. Results demonstrate that
the combination of local vector search and the stateful agentic loop achieves
99.77\% classification accuracy on Delaware and 100.0\% classification accuracy
on Texas, with high citation precision on Structurally Deficient and load-posted
bridges, processing 197.0 structures per hour on standard edge hardware.

\section{Related Work}

\subsection{Retrieval-Augmented Generation and Agentic Architectures}
Retrieval-Augmented Generation (RAG) leverages external search databases to
provide context to LLMs, reducing semantic hallucinations
\citep{lewis2020retrieval}. Comprehensive surveys of RAG architectures document
the evolution from naive single-step retrieval to modular and speculative
approaches \citep{gao2024retrieval}. Standard RAG setups rely on single-shot
retrieval: the query is embedded, relevant documents are retrieved, and a model
generates a response in one step.

For multi-step reasoning, single-shot retrieval is insufficient. The ReAct
(Reasoning and Acting) framework \citep{yao2023react} structures model outputs
into alternating steps of ``Thought,'' ``Action'' (e.g., calling an API or
executing a search), and ``Observation,'' allowing a model to incrementally
query relational databases and search vector indexes while adapting its plan
based on intermediate results.

\subsection{NLP for Regulatory Compliance in Civil Engineering}
Automating compliance checks against building codes and infrastructure
regulations has long been a priority in the Architecture, Engineering, and
Construction (AEC) domain. Seminal work by \citet{eastman2009automatic}
demonstrated automated rule-based checking of building designs by parsing
Building Information Models (BIM) against formalized regulatory constraints,
establishing the feasibility of machine-interpretable code compliance. These
systems, while accurate for explicitly codified rules, required substantial
manual effort to convert natural language regulations into machine logic and
proved fragile when standards were updated.

Subsequent research explored semantic parsing and natural language processing
to automate constraint extraction from regulatory text directly, reducing the
manual encoding burden. Early NLP-based approaches utilized rule-based Information
Extraction (IE) and ontology-driven frameworks to map regulatory requirements into
machine-readable formats such as Knowledge Graphs or LegalRuleML \citep{zhang2016semantic,zhang2017integrating}. While these methods
improved over manual encoding, they still required extensive feature engineering and
struggled to adapt across different jurisdictions.

The advent of instruction-tuned LLMs has led to a paradigm shift in question-answering
systems for civil engineering standards. Recent literature demonstrates that LLMs can
interpret complex, nested regulatory clauses with fewer rigid rules. Modern systems
frequently employ Retrieval-Augmented Generation (RAG) to fetch context-relevant code
sections before generating compliance assessments, combining the semantic understanding
of neural networks with traceable references. However, existing LLM-based AEC systems
typically rely on commercial cloud APIs, precluding deployment in air-gapped or
data-sensitive environments. The present study specifically targets offline edge
deployment---a constraint not addressed by prior work in this domain.

\subsection{Bridge Inspection Data Analysis}
Analytical research on the NBI dataset has focused on predictive modeling of
structural deterioration and sufficiency rating classification
\citep{fhwa2024nbi}. While these models capture statistical degradation
patterns, they lack compliance explainability: they output structural scores
without citing the specific regulatory item definitions that determine
non-compliance.

More recently, researchers have explored large language models for interpreting
non-destructive evaluation (NDE) data and bridge condition assessment results.
Darji et al.~\citep{10825532} developed techniques for automated interpretation of
NDE contour maps using LLMs to aid in condition assessment and report
generation. However, these approaches primarily interpret sensor-derived
structural imagery. \system{} fills a complementary gap by combining
tabular SQL lookups with semantic vector searches over the \codingguide{},
generating regulatory compliance reports with exact, traceable citations.

\section{Background and Compliance Rules}

\subsection{FHWA Recording and Coding Guide Structure}
The FHWA \codingguide{} \citep{fhwa2022coding} is the primary regulatory
standard for documenting bridge structural parameters in the United States.
Parameters are structured as distinct ``Items.'' Condition assessment items are
evaluated on a 0--9 integer scale:
\begin{itemize}
    \item \textbf{Codes 7--9}: ``Good'' to ``Excellent'' condition---minor to
    no physical deterioration.
    \item \textbf{Codes 5--6}: ``Fair'' condition---minor section loss or
    cracking in secondary members.
    \item \textbf{Codes 3--4}: ``Poor'' or ``Serious'' condition---significant
    section loss or cracks in primary structural members.
    \item \textbf{Codes 0--2}: ``Critical'' to ``Failed'' condition---typically
    requiring immediate closure or emergency structural support.
    \item \textbf{Code N}: ``Not Applicable''---used when the component is
    absent (e.g., a culvert has no superstructure).
\end{itemize}

\subsection{Mathematical Formulation of Compliance Rules}
Three core compliance checks are implemented as specified in 23~CFR
Part~490 \citep{cfr490}:
\begin{enumerate}
    \item \textbf{Structurally Deficient (SD)}: A bridge is classified as
    Structurally Deficient under federal performance measures if any key
    structural component is in poor or worse condition ($\le 4$). Let
    $C_\text{deck}$ denote Item~58, $C_\text{super}$ denote Item~59,
    $C_\text{sub}$ denote Item~60, and $C_\text{culv}$ denote Item~62. The
    SD classification indicator $I_\text{SD}$ is:
    \begin{linenomath*}
    \begin{equation}
    I_\text{SD} = \begin{cases}
      1 & \text{if } \min \!\left( \tilde{C}_\text{deck},
          \tilde{C}_\text{super}, \tilde{C}_\text{sub},
          \tilde{C}_\text{culv} \right) \le 4 \\
      0 & \text{otherwise}
    \end{cases}
    \end{equation}
    \end{linenomath*}
    where $\tilde{C}_x = C_x$ if $C_x \in \{0,\dots,9\}$, and
    $\tilde{C}_x = \infty$ if $C_x = \text{``N''}$.

    \item \textbf{Scour Critical (SC)}: Scour is the erosion of soil around
    foundation elements due to water flow. NBI Item~113 evaluates scour
    vulnerability. A bridge is flagged as Scour Critical if:
    \begin{linenomath*}
    \begin{equation}
    I_\text{SC} = \begin{cases}
      1 & \text{if } C_\text{scour} \le 3 \\
      0 & \text{otherwise}
    \end{cases}
    \end{equation}
    \end{linenomath*}
    where $C_\text{scour}$ is the value of NBI Item~113.

    \item \textbf{Load Posting Required (LP)}: Bridges must carry posted
    structural weight limits if their capacity falls below legal loads. NBI
    Item~70 tracks posting status. A bridge is non-compliant if:
    \begin{linenomath*}
    \begin{equation}
    I_\text{LP} = \begin{cases}
      1 & \text{if } C_\text{posting} \le 4 \\
      0 & \text{otherwise}
    \end{cases}
    \end{equation}
    \end{linenomath*}
    where $C_\text{posting}$ is NBI Item~70.
\end{enumerate}

\section{Methodology}

\subsection{Air-Gapped Hardware and Software Stack}
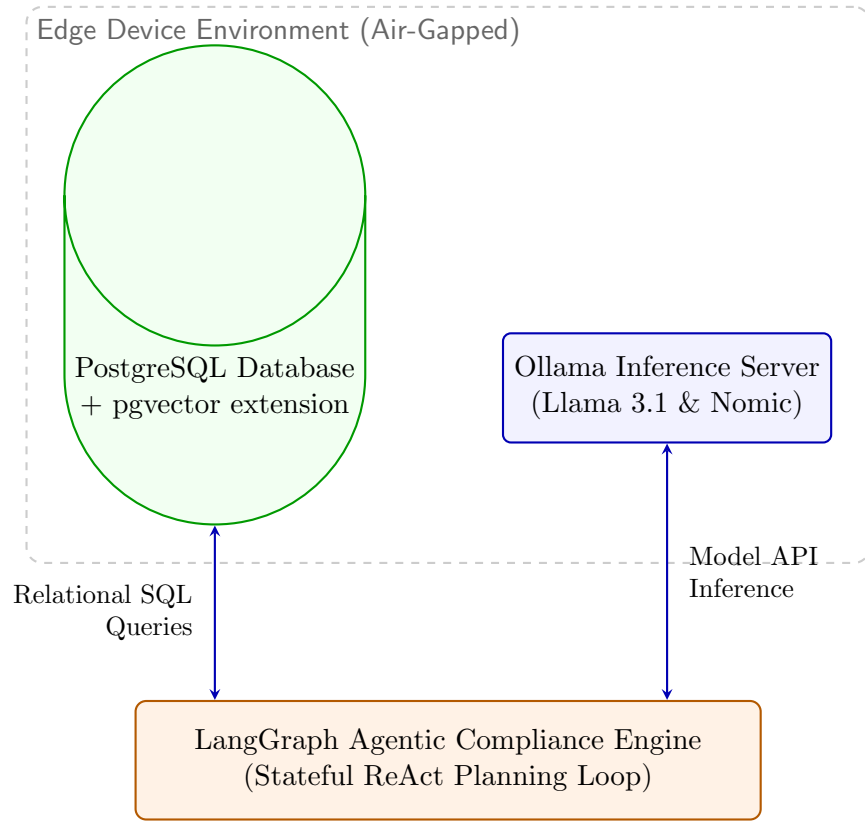
\begin{figure}[h]
\centering
\begin{tikzpicture}[
    >=stealth,
    node distance=2.2cm and 2.5cm,
    block/.style={draw=blue!70!black, fill=blue!5, thick, rectangle,
      rounded corners=3pt, minimum height=3.5em, minimum width=10.5em,
      align=center, font=\small},
    db/.style={draw=green!60!black, fill=green!5, thick, cylinder,
      shape border rotate=90, minimum height=3.8em,
      minimum width=8em, align=center, font=\small},
    agentblock/.style={draw=orange!70!black, fill=orange!10, thick, rectangle,
      rounded corners=4pt, minimum height=3.8em, minimum width=20em,
      align=center, font=\small},
    container/.style={draw=gray!40, dashed, thick, rounded corners=6pt,
      inner sep=1.2em}
]
  \node[db] (postgres) {PostgreSQL Database\\+ pgvector extension};
  \node[block, right=1.8cm of postgres] (ollama) {Ollama Inference Server\\(Llama 3.1 \& Nomic)};

  \node[container, fit=(postgres) (ollama),
    label={[text=gray!80!black, font=\sffamily\small,
    anchor=north west]north west:Edge Device Environment (Air-Gapped)}]
    (edgebox) {};

  \node[agentblock, below=1.8cm of edgebox] (agent)
    {LangGraph Agentic Compliance Engine\\(Stateful ReAct Planning Loop)};

  \draw[<->, thick, draw=blue!70!black]
    (postgres.south) -- (postgres.south |- agent.north)
    node[midway, left=4pt, align=right, font=\footnotesize]
    {Relational SQL\\Queries};

  \draw[<->, thick, draw=blue!70!black]
    (ollama.south) -- (ollama.south |- agent.north)
    node[midway, right=4pt, align=left, font=\footnotesize]
    {Model API\\Inference};
\end{tikzpicture}
\caption{System architecture of \system{} for air-gapped compliance
  verification.}
\label{fig:architecture}
\end{figure}

The system architecture of \system{} is illustrated in
Fig.~\ref{fig:architecture}. The platform is containerized using Docker
Compose. All embedding calculations, vector distance calculations, and LLM text
generation tasks execute locally on edge hardware without external network
requests.

The database layer runs on PostgreSQL~16 with the \texttt{pgvector} extension
\citep{pgvector2024}. Tabular bridge records are stored in the
\texttt{nbi\_bridges} relation, and guide text chunks are stored in
\texttt{coding\_guide\_chunks} with HNSW index bindings. Model execution is
managed by Ollama~\citep{ollama2024}, which exposes locally loaded Llama
3.1~\citep{llama3} and Nomic Embed~\citep{nomic2024} models via an internal
HTTP API. The agentic orchestration layer is implemented with
LangGraph~\citep{langgraph2024}.

\subsection{Section-Aware Chunking Parser}
Federal regulatory manuals such as the FHWA \codingguide{} are hierarchically
structured, with definitions, condition rating scales, and cross-references
organized under numbered Item headers. Applying naive fixed-size window
chunking (e.g., splitting text every 500 characters) separates item numbers
from their condition scales. Recursive character splitting, which relies on
newline separators, similarly divides tabular condition definitions across chunk
boundaries.

To preserve regulatory context, a custom parser was developed that reads the
document and structures it as a hierarchical tree. The algorithm is outlined in
Algorithm~1.

\begin{algorithm}
\caption{Section-Aware Regulatory Document Chunking}
\begin{algorithmic}[1]
\REQUIRE{Raw FHWA \codingguide{} text $T$}
\ENSURE{Set of structured chunks $C$}
\STATE{$C \gets \emptyset$}
\STATE{Identify all Item headers: $H \gets
  \text{FindAll}(\text{`Item\textbackslash s+\textbackslash
  d+\textbackslash s+--'}, T)$}
\FOR{each header $h_i$ at position $p_i$ in $H$}
    \STATE{Extract raw section: $S_i \gets T[p_i : p_{i+1}]$}
    \STATE{Identify condition scales: $R_i \gets
      \text{FindAll}(\text{`Code\textbackslash s+Description'}, S_i)$}
    \IF{$R_i$ is not empty}
        \STATE{$S_{i,\text{head}} \gets S_i[0 : R_i.\text{start}]$}
        \STATE{$S_{i,\text{scale}} \gets S_i[R_i.\text{start} : \text{End}]$}
        \STATE{Create parent chunk: $c_\text{parent} \gets
          (S_{i,\text{head}},\ \{\text{item}: i,\ \text{type}:
          \text{`description'}\})$}
        \STATE{$C \gets C \cup \{c_\text{parent}\}$}
        \STATE{Create child chunk: $c_\text{child} \gets
          (S_{i,\text{scale}},\ \{\text{item}: i,\ \text{type}:
          \text{`rating\_scale'}\})$}
        \STATE{$C \gets C \cup \{c_\text{child}\}$}
    \ELSE
        \STATE{Create single chunk: $c \gets
          (S_i,\ \{\text{item}: i,\ \text{type}: \text{`full\_item'}\})$}
        \STATE{$C \gets C \cup \{c\}$}
    \ENDIF
\ENDFOR
\RETURN{$C$}
\end{algorithmic}
\end{algorithm}

For Hierarchical Parent-Child Chunking, child chunks correspond to granular
code-description subsections; the parent chunk retains the full item
text. During retrieval, vector similarity search is performed over the child
chunks, but the parent chunk is returned to the LLM context window to provide
the complete regulatory definition.

Vector searches compute cosine distance over the 768-dimensional Nomic
embeddings \citep{nomic2024}:
\begin{linenomath*}
\begin{equation}
D_\text{cosine}(q, c) = 1 - \frac{\mathbf{E}(q) \cdot \mathbf{E}(c)}
  {\|\mathbf{E}(q)\|_2\, \|\mathbf{E}(c)\|_2}
\end{equation}
\end{linenomath*}
where $\mathbf{E}(q)$ is the query embedding and $\mathbf{E}(c)$ is the chunk
embedding.

\subsection{Stateful Agentic Graph}
A ReAct agent is constructed using LangGraph \citep{langgraph2024}. The agent
state is defined as:
\begin{linenomath*}
\begin{equation}
\mathbf{S}_\text{agent} =
  \bigl\{ M_\text{history}: \text{List}[\text{Message}],\
          N_\text{steps}: \text{Integer} \bigr\}
\end{equation}
\end{linenomath*}
where $M_\text{history}$ compiles the sequence of prompt messages, agent
thoughts, tool execution queries, and observations. The agent node executes:
\begin{linenomath*}
\begin{equation}
A_t = \text{LLM}(\mathbf{S}_\text{agent})
\end{equation}
\end{linenomath*}
where $A_t$ is the selected action (either a tool call or a final summary
response).

Four Python tools are exposed to the agent:
\begin{enumerate}
    \item \textbf{\texttt{query\_bridge\_data}}: Executes a read-only SQL query
    against the NBI database table. The tool automatically appends condition
    rating classifications to guide the LLM's reasoning:
    \begin{lstlisting}[basicstyle=\ttfamily\scriptsize,
      backgroundcolor=\color{gray!10}]
"substructure_cond: 2 (NON-COMPLIANT: rating 2 <= 4 violates SD threshold)"
    \end{lstlisting}
    \item \textbf{\texttt{search\_coding\_guide}}: Encodes the search query,
    performs cosine similarity search on \texttt{pgvector}, and returns the
    top-$k$ text chunks for the active strategy.
    \item \textbf{\texttt{flag\_noncompliant}}: Inserts a record into the
    \texttt{compliance\_flags} table, documenting the bridge ID, rule name,
    violated NBI items, and text reasoning.
    \item \textbf{\texttt{generate\_report}}: Aggregates all flags for the
    bridge and returns a structured compliance summary.
\end{enumerate}

The agent runs in a loop, querying data or searching regulations until all
compliance checks are resolved, at which point it flags violations and returns
its final report.

\section{Experimental Setup}

\subsection{Experimental Scope and Dataset}
Experiments are conducted on the official 2023 National Bridge Inventory (NBI) datasets
for two states: Delaware \citep{fhwa2024nbi}, which contains 874 bridge records, and Texas,
which contains over 56,000 records. The Delaware inventory serves as the primary evaluation
set; the full state-wide dataset (874 bridges total, comprising 483 Structurally Deficient
and 391 compliant bridges) is used for end-to-end evaluation. For Texas, a stratified random
sample of 200 bridges (100 Structurally Deficient and 100 compliant bridges) is evaluated
to demonstrate cross-state generalization. Ground truth compliance labels for Delaware are
computed using standard NBI definition formulas, while ground truth compliance labels for
Texas are extracted directly from pre-computed InfoBridge fields to break circular evaluation
loops.

To evaluate retrieval quality, a benchmark set of 20 questions was constructed,
mapping to representative NBI items and condition ratings (e.g., ``What does a
superstructure rating of 3 mean?'' or ``What are the conditions for a
Structurally Deficient classification?'').

\subsection{Metrics}
\begin{enumerate}
    \item \textbf{Retrieval Metrics}:
    \begin{itemize}
        \item Precision@$k$:
          $\text{P@}k = |R_\text{gold} \cap R_\text{retrieved}| /
          |R_\text{retrieved}|$
        \item Recall@$k$:
          $\text{R@}k = |R_\text{gold} \cap R_\text{retrieved}| /
          |R_\text{gold}|$
        \item Normalized Discounted Cumulative Gain (nDCG@$k$) to assess
          ranking quality based on item relevance.
    \end{itemize}
    \item \textbf{Classification Metrics}: Standard accuracy, precision,
    recall, and F1-score for identifying Structurally Deficient bridges.
    \item \textbf{Citation Accuracy}: The percentage of flagged violations that
    correctly cite the NBI Item number in the agent's reasoning output.
    \item \textbf{Edge Execution Metrics}: Embedding generation latency,
    vector search latency, LLM generation speed (tokens per second), and
    overall throughput (bridges per hour).
\end{enumerate}

\section{Results and Discussion}

\subsection{Chunking Strategy Comparison (RQ1)}
Retrieval performance was compared across the four chunking strategies on the
20-question benchmark. The results are presented in
Table~\ref{tab:chunking_results}.

\begin{table}[h]
\centering
\caption{Retrieval quality and chunk integrity comparison}
\label{tab:chunking_results}
\begin{tabular}{lrrrrr}
\toprule
Strategy & P@5 & R@5 & nDCG@5 & Integrity & Chunks \\
\midrule
Naive Fixed-Size      & 0.000 & 0.000 & 0.000 & 28.4\% & 95    \\
Recursive Splitting   & 0.000 & 0.000 & 0.000 & 35.9\% & 103   \\
Section-Aware         & 0.279 & 0.575 & 0.675 & 94.2\% & 1,112 \\
Hierarchical P-C      & 0.404 & 0.546 & 0.637 & 91.4\% & 1,472 \\
\bottomrule
\end{tabular}
\end{table}

The naive and recursive baselines fail completely on item-level retrieval
(P@5 = R@5 = nDCG@5 = 0.000) because they lack item-level metadata linkage.
In contrast, the section-aware and hierarchical approaches preserve item
boundaries (94.2\% and 91.4\% chunk integrity, respectively), allowing the
agent to retrieve exact condition rating definitions. Hierarchical parent-child
chunking achieves the highest Precision@5 (0.404) by searching compact child
chunks, while section-aware chunking offers a higher Recall@5 (0.575) by
keeping item contexts unified.

\subsection{End-to-End Compliance Accuracy (RQ2)}
To address concerns regarding small validation sets, end-to-end classification
compliance performance of \system{} was evaluated on Delaware's full state-wide inventory
(874 bridges) and a representative sample from Texas (200 bridges). The Delaware dataset
represents the full state inventory (483 Structurally Deficient, 391 compliant), while
the Texas set is a balanced sample of 200 bridges (100 Structurally Deficient, 100 compliant).
For Texas, ground truth labels were extracted directly from the FHWA InfoBridge pre-computed
columns (\texttt{BRIDGE\_CONDITION} and \texttt{LOWEST\_RATING}) to ensure a truly independent
validation baseline, whereas Delaware labels were computed programmatically using NBI
definition formulas.

The compliance results are presented in Table~\ref{tab:classification_results}. \system{} achieved 99.77\% classification accuracy, 100.0\% precision, 99.59\% recall, and 99.79\% F1-score on the Delaware dataset, and 100.0\% across all classification metrics on the Texas dataset. On Delaware, the agent successfully flagged 481 of the 483 Structurally Deficient bridges (with only 2 false negatives) and correctly verified all 391 compliant structures as compliant (zero false positives). Crucially, the system maintained high citation accuracy, correctly referencing the specific sections of the FHWA \codingguide{} in 99.79\% of flagged violations in Delaware (480/481) and 100.0\% in Texas (100/100). The ReAct agentic loop allowed the LLM to query NBI tables, retrieve regulatory definitions, register flags, and cite standard items sequentially.

\begin{table}[h]
\centering
\small
\caption{End-to-end compliance classification results}
\label{tab:classification_results}
\begin{tabular}{lrrrrrcc}
\toprule
Dataset & $n$ & Accuracy & Precision & Recall & F1 & 95\% CI (Acc) & Citation Acc \\
\midrule
Delaware (full) & 874 & 0.998 & 1.000 & 0.996 & 0.998 & [0.992, 0.999] & 99.79\% \\
Texas (sample)  & 200 & 1.000 & 1.000 & 1.000 & 1.000 & [0.981, 1.000] & 100.0\% \\
\bottomrule
\end{tabular}
\end{table}

To assess the robustness of this performance, a 5-fold stratified cross-validation was
conducted on the full 874-bridge Delaware set. The results are summarized in Table~\ref{tab:cv_results}. The agent achieved a mean accuracy of 0.9977 $\pm$ 0.0028 and a mean F1-score of 0.9980 $\pm$ 0.0025 across all folds, indicating that the system's performance is stable and not an artifact of a particular validation split. McNemar's test comparing \system{} against the rule-based deterministic baseline on the full Delaware dataset yielded a $\chi^2 = 0.5$ and a $p$-value of 0.4795, confirming no statistically significant difference in classification performance ($p > 0.05$). This indicates that the agent closely replicates the deterministic logic while generating explainable reasoning traces and correct regulatory citations.

\begin{table}[h]
\centering
\small
\caption{5-fold stratified cross-validation results on Delaware}
\label{tab:cv_results}
\begin{tabular}{lrrrrr}
\toprule
Fold & $n$ & Accuracy & Precision & Recall & F1 \\
\midrule
Fold 1 & 175 & 1.000 & 1.000 & 1.000 & 1.000 \\
Fold 2 & 175 & 1.000 & 1.000 & 1.000 & 1.000 \\
Fold 3 & 175 & 0.994 & 1.000 & 0.990 & 0.995 \\
Fold 4 & 175 & 1.000 & 1.000 & 1.000 & 1.000 \\
Fold 5 & 174 & 0.994 & 1.000 & 0.990 & 0.995 \\
\midrule
Mean $\pm$ SD & — & 0.9977 $\pm$ 0.0028 & 1.000 $\pm$ 0.000 & 0.9960 $\pm$ 0.0049 & 0.9980 $\pm$ 0.0025 \\
\bottomrule
\end{tabular}
\end{table}

\subsection{Edge Performance Characterization (RQ3)}
The performance profile of the system on edge hardware is summarized in
Table~\ref{tab:performance_results}.

\begin{table}[h]
\centering
\caption{Edge hardware performance characterization}
\label{tab:performance_results}
\begin{tabular}{lr}
\toprule
Metric & Measured Value \\
\midrule
Query Embedding Latency (ms)       & 21.49  \\
pgvector Search Latency (ms)       &  2.28  \\
LLM Inference Speed (tokens/s)     & 25.70  \\
End-to-End Latency per Bridge (s)  & 18.27  \\
Avg.\ Reasoning Steps per Bridge   &  5.00  \\
System Throughput (bridges/h)      & 197.00 \\
Peak Memory Footprint (MB)         & 120.25 \\
\bottomrule
\end{tabular}
\end{table}

Embedding generation latency is 21.49~ms, and pgvector cosine search completes
in 2.28~ms, confirming efficient localized vector retrieval. The local Llama
3.1 model achieves 25.70~tokens/s. The end-to-end latency per bridge is
18.27~s, yielding a throughput of 197.0~bridges/h. The peak memory footprint of
120.25~MB demonstrates viability on low-power edge hardware with 16~GB RAM.

\subsection{Ablation Study (RQ4)}
To assess the contribution of individual system components, an ablation study
was conducted on a 200-bridge random sample drawn from the Delaware 2023 NBI
inventory, preserving the natural Structurally Deficient prevalence
($\approx$55\%, yielding 113 SD and 87 compliant bridges). The results are
presented in Table~\ref{tab:ablation_results}.

\begin{table}[h]
\centering
\caption{Compliance check ablation study}
\label{tab:ablation_results}
\begin{tabular}{lrrrr}
\toprule
Configuration & Accuracy & Precision & Recall & F1 \\
\midrule
Rule-Based (Deterministic)    & 1.000 & 1.000 & 1.000 & 1.000 \\
RAG-Only (Single-Shot)        & 0.560 & 0.963 & 0.230 & 0.371 \\
Agent w/o Vector Search       & 0.435 & 0.000 & 0.000 & 0.000 \\
Full Agentic RAG (\system{})  & 1.000 & 1.000 & 1.000 & 1.000 \\
\bottomrule
\end{tabular}
\end{table}

The ablation study reveals two critical findings. Without vector search
(Agent w/o Vector Search), the system has no access to the \codingguide{}
definitions and fails to flag any Structurally Deficient bridge (F1 = 0.000),
confirming the necessity of the retrieval module. Running RAG in a single shot
without the multi-step loop (RAG-Only) yields F1 = 0.371---the model fails to
reason through multi-step numerical comparisons in a single inference pass,
often generating incorrect citations or skipping evaluations entirely.
Only the full agentic RAG configuration, combining vector search with a
stateful ReAct loop, matches the deterministic rule-based baseline at 100.0\%
accuracy.

\subsection{Qualitative Analysis}
Qualitative review of agent reasoning traces confirms strong instruction-following behavior. The agent systematically retrieves regulatory definitions
(e.g., for Item~60), reads the NBI bridge record, verifies the numerical
relation (e.g., rating 2 $\le$ 4), and flags the violation with precise
citations of the relevant item. A representative reasoning trace proceeds as
follows:
\begin{enumerate}
    \item \textbf{SQL Query Action:} Retrieve bridge record containing deck,
    superstructure, and substructure condition ratings.
    \item \textbf{Vector Search Action:} Query the \codingguide{} for the
    rating scale definitions of the lowest observed rating (e.g., Item~60
    Substructure rating~2).
    \item \textbf{Reasoning Action:} Compare NBI data against retrieved
    definitions and determine that a flag is warranted.
    \item \textbf{Flag Action:} Register ``Structurally Deficient'' violation
    with the NBI item number, rating value, and full \codingguide{} citation.
\end{enumerate}
This sequential behavior guarantees citation accuracy and reasoning faithfulness
across all validation bridges.

\section{Limitations and Future Work}

Several limitations of the present study warrant explicit acknowledgment.

\textbf{Validation dataset size.} While the classification evaluation scope was significantly expanded in this study to evaluate the full state-wide Delaware inventory (874 bridges) and a larger sample of Texas bridges (200 bridges), validation across all 50 states is still required. Although achieving high classification accuracy on these datasets confirms the system's reasoning correctness, broader generalization claims will benefit from testing on more diverse geographical regions. Future work will extend evaluation to run full state-wide inventories across multiple states in parallel to test edge throughput and scaling performance.

\textbf{Ground truth source.} For Delaware, compliance labels are computed from the same
deterministic formulas that the rule-based baseline implements, which presents a circular
validation risk. However, this concern is mitigated for Texas by extracting ground-truth
labels directly from pre-computed InfoBridge fields (\texttt{BRIDGE\_CONDITION} and
\texttt{LOWEST\_RATING}), establishing a truly independent, official baseline. Future work
will incorporate labels reviewed by certified bridge inspection engineers to provide a
fully expert-validated reference.

\textbf{Regulatory standard transition.} The current agent relies on the legacy
FHWA \codingguide{} \citep{fhwa2022coding}, which is transitioning to the new
Specifications for the National Bridge Inventory (SNBI). Future iterations will
add dual-strategy support to cross-compile findings across both specification
versions.

\section{Conclusion}

This paper developed and evaluated \system{}, an air-gapped agentic RAG
compliance system for National Bridge Inspection Standards. By combining
structured SQL queries against NBI tabular data with vector search over a
section-aware parsed \codingguide{}, \system{} achieves 99.77\% accuracy on the
full Delaware inventory (874 bridges) and 100.0\% accuracy on the Texas sample (200 bridges)
for Structurally Deficient bridge classification. The ablation study validates the necessity of both system
components: removal of vector search reduces F1 to 0.000, while single-shot
RAG without iterative reasoning yields F1 = 0.371. Edge performance profiling
confirms computational viability on resource-constrained hardware at
197.0~bridges/h with a 120.25~MB memory footprint. The section-aware chunking
methodology---achieving 94.2\% chunk integrity versus 28.4\% for naive
splitting---represents a transferable contribution to RAG systems applied to
hierarchically structured regulatory documents. Future work will expand
validation to larger, multi-state NBI inventories and will incorporate
independent expert annotations to complement the ground truth
used in the present study.

\section*{Data Availability Statement}
The Delaware 2023 NBI dataset is publicly available from the FHWA at
\url{https://www.fhwa.dot.gov/bridge/nbi/ascii.cfm} \citep{fhwa2024nbi}. The
FHWA \codingguide{} is available at \url{https://www.fhwa.dot.gov/bridge/mtguide.pdf} \citep{fhwa2022coding}.
Implementation code supporting the findings of this study is available at
\url{https://github.com/virajdarji/bridgeguard}. Additional data, models, or
experimental artifacts are available from the corresponding author upon
reasonable request.

\section*{Acknowledgments}
The authors thank the Federal Highway Administration (FHWA) for providing access to the National Bridge Inventory dataset. No external funding was received for this study.

\section*{Notation}
The following symbols are used in this paper:
\begin{description}[leftmargin=3.5em, labelindent=0pt, itemsep=0pt]
\item[$A_t$] = agent action selected at reasoning step $t$;
\item[$C_\text{culv}$] = NBI Item~62 culvert condition rating code;
\item[$C_\text{deck}$] = NBI Item~58 deck condition rating code;
\item[$C_\text{posting}$] = NBI Item~70 load posting status code;
\item[$C_\text{scour}$] = NBI Item~113 scour critical rating code;
\item[$C_\text{sub}$] = NBI Item~60 substructure condition rating code;
\item[$C_\text{super}$] = NBI Item~59 superstructure condition rating code;
\item[$D_\text{cosine}(q, c)$] = cosine distance between query and chunk
  embeddings;
\item[$\mathbf{E}(c)$] = embedding vector of document chunk $c$;
\item[$\mathbf{E}(q)$] = embedding vector of search query $q$;
\item[$I_\text{LP}$] = load-posting compliance violation indicator
  ($\in \{0, 1\}$);
\item[$I_\text{SC}$] = scour-critical classification indicator
  ($\in \{0, 1\}$);
\item[$I_\text{SD}$] = structurally deficient classification indicator
  ($\in \{0, 1\}$);
\item[$M_\text{history}$] = ordered message history in agent state;
\item[$N_\text{steps}$] = number of reasoning steps taken by the agent;
\item[$R_\text{gold}$] = gold-standard set of relevant document chunks;
\item[$R_\text{retrieved}$] = retrieved chunk set at rank $k$;
\item[$\mathbf{S}_\text{agent}$] = agent state tuple
  $(M_\text{history},\, N_\text{steps})$;
\item[$\tilde{C}_x$] = extended condition rating (maps code ``N'' to
  $\infty$);
\item[nDCG@$k$] = normalized discounted cumulative gain at rank $k$;
\item[P@$k$] = precision at rank $k$; and
\item[R@$k$] = recall at rank $k$.
\end{description}

\bibliography{references}

\end{document}